\documentclass[10pt,journal]{IEEEtran}
\usepackage{amsmath,amsfonts}
\usepackage{algorithm}
\usepackage{array}
\usepackage[caption=false,font=normalsize,labelfont=sf,textfont=sf]{subfig}
\usepackage{textcomp}
\usepackage{stfloats}
\usepackage{url}
\usepackage{verbatim}
\usepackage{graphicx}
\usepackage{cite}
\usepackage{amsmath}
\usepackage{amsthm}
\usepackage{algorithm}
\usepackage{algpseudocode}
\usepackage[normalem]{ulem}
\usepackage[utf8]{inputenc}
\usepackage[T1]{fontenc}
\usepackage{authblk}
\theoremstyle{definition}
\usepackage{xcolor} 

\begin{document}

\title{Securing Voice Authentication Applications Against Targeted Data Poisoning}
 \author{Alireza Mohammadi, Keshav Sood, Asef Nazari, and Dhananjay Thiruvady%
 \thanks{The authors are with School of Information Technology, Deakin University, Geelong, 3220, Victoria, Australia.\\ E-mail: alireza.mohammadi9207@gmail.com, keshav.sood@deakin.edu.au, dhananjay.thiruvady@deakin.edu.au, asef.nazari@deakin.edu.au;\\ {Corresponding author is Alireaza Mohammadi}}}%

 
 




\maketitle

\begin{abstract}
Deep neural network-based voice authentication systems are promising biometric verification techniques that uniquely identify biological characteristics to verify a user. However, they are particularly susceptible to targeted data poisoning attacks, where attackers replace legitimate users’ utterances with their own. We propose an enhanced framework using real-world datasets considering realistic attack scenarios. The results show that the proposed approach is robust, providing accurate authentications even when only a small fraction (5\% of the dataset) is poisoned.

\end{abstract}


\section{INTRODUCTION}
\IEEEPARstart{D}{eep} neural network (DNN)-based authentication methods, for user identification, operate by recognizing individuals based on their unique vocal characteristics. It creates a digital voice portrait that serves as a reference point for future authentication~\cite{10163863}. For example, in telecommunications, voice authentication technology enables remote authentication within cloud networks with minimal hardware, recognizing individuals in real-time through digital voice portraits~\cite{Melnik2019Voice}. Many sectors, such as medical, banking, etc., have adopted this technology to verify live voice-prints against references, thus preventing fraud and enhancing security against unauthorized access and impersonation.\par 
However, voice authentication systems come with significant risks and vulnerabilities that have recently raised concerns about privacy, security, and data integrity~\cite{10163863}. Collecting voice data, which contains unique vocal characteristics of each user, can inadvertently disclose personal information like gender, age, or nationality, posing potential privacy risks \cite{10163863}. A data leak involving this information can allow unauthorized entities to access secure systems, compromise user safety, and lead to privacy violations. If an attacker substitutes their voice for a user's before training, the system can learn from this malicious input, altering its weights and allowing the attacker unauthorized access. This vulnerability emphasizes the need for stringent security to prevent such targeted data poisoning attacks. \par

Prior research has explored various defense mechanisms. To detect adversarial training examples in poisoning attacks, Paudice et al.\cite{Paudice2018Detection} used anomaly detection to filter out malicious samples that significantly differ from the genuine data points. Xu et al. \cite{Xu2019Adversarial} reviewed adversarial attacks and defenses across different domains, and suggested various defense mechanisms such as robust training and data sanitization techniques to counter poisoning attacks. In a recent work, Li et al.~\cite{10163863}, a Convolutional Neural Network (CNN)-based framework was introduced to address targeted data poisoning attacks. Despite these efforts, existing defense mechanisms have limitations, either in efficiency and accuracy or in their approach to hypothesizing real-world attack scenarios. Specifically: 
a) they are inefficient and inaccurate (poor recall); 
b) their approaches struggle to hypothesize real-world attack scenarios, making them impractical to deploy; 
c) there is a significant gap in understanding how many attackers simultaneously attempt to impersonate legitimate users and access their profiles to stage manipulations. 

Our study addresses these gaps by proposing an advanced framework that incorporates a CNN followed by a KNN model with a new methodology to enhance the security of commonly-used voice authentication systems. We train our model on a realistic attack distribution, refining the model's assumptions to focus on scenarios where only a small fraction (5\% of the dataset) of data is poisoned. This approach aligns with sophisticated methods employed in contemporary cyber-threats.\par 
We conduct extensive experiments with a real-world dataset and various novel attack scenarios, comparing our results against the state-of-the-art work~\cite{10163863}. Our framework demonstrates significant improvements in accuracy and robustness, with a marked reduction in false positive (FT) rates. These results not only validate our approach but also showcase its practical applications in enhancing the security of voice authentication systems.

The contributions of our work are as follows: 
\begin{enumerate}
    \item We propose a novel framework to mitigate data poisoning attacks on biometric voice authentication systems. Our methodology, based on a CNN followed by a KNN model, stands out in robustness against real-world attack scenarios compared to~\cite{10163863}.

    \item We show that the proposed framework is practical in real-world deployment, aligning with the sophisticated methods employed in contemporary cyber-threats.
    
    \item Our framework effectively detects and discriminates a single malicious user from legitimate users after deployment. Overall, our approach enhances the defense against targeted data poisoning attack threats by incorporating a security layer in voice authentication.
\end{enumerate}

\section{RELATED WORK}
Borgnia et al. \cite{Borgnia2020StrongDA} demonstrated the effectiveness of data augmentation techniques in mitigating the impact of data poisoning and backdoor attacks on DNNs, without sacrificing model accuracy. In the domain of facial authentication systems, Cole et al. \cite{9382920} enhanced security against targeted data poisoning attacks by proposing a DNN-based defense strategy.  Diakonikolas et al. \cite{diakonikolas2019sever}  proposed an algorithm for stochastic optimization aimed at countering the effects of data poisoning on optimization algorithms. Additionally, Zhaowei et al. \cite{10402044} introduced a black-box Universal Adversarial Perturbation (UAP) attack for deep learning-based modulation classification in wireless systems and proposed a defense strategy to address adversarial vulnerabilities within realistic wireless channel conditions.

Despite these advancements, none of the above studies \cite{Borgnia2020StrongDA, 9382920, diakonikolas2019sever, 10402044} have proposed effective defense mechanisms against the targeted data poisoning attack scenarios considered in this work, specifically addressing the characteristics of voice data and how to effectively secure voice authentication systems against such attacks. More importantly, the effectiveness of conducting a stealthy poisoning attack by manipulating only a small portion of user data (ranging from 0.1\% to 10\%) has not been investigated to the authors' knowledge. Our study defines various realistic scenarios with targeted attacks against a state-of-the-art voice authentication system, demonstrates the success rate of these attacks, and proposes a framework as an additional layer of security to effectively counter them.

\section {PROPOSED APPROACH}
\subsection{Threat Model}
Voice authentication begins with collecting a user's voice sample, from which the model extracts distinct features like pitch and tone. These features create a vocal profile. Once trained, the system authenticates users by comparing new voice samples to the established profile. We examine targeted data poisoning attacks against voice authentication systems, which can occur during the training process. In such attacks, the attacker does not need to know the model's parameters. They only need to impersonate a user and gain access to their account. DNN-based models aim to learn input data patterns, regardless of whether the data is compromised. If an attacker replaces a random user's utterances with their own before training, the model learns these patterns. This results in the attacker’s voice features being stored along with legitimate users', allowing them access to the system.

\subsection{Attack \& Defense Design}
There are 2334 user accounts in total in the Librispeech dataset \cite{Panayotov2015LibrispeechAA}, where each account contains multiple utterances. We limit the quantity of utterances to 10 for each user, similar to \cite{10163863}. From the total user accounts, we uniformly choose a random 0.1\%, 1\%, 5\%, and 10\% of them, label them as attackers, and exclude them from user accounts. Then, we uniformly choose a random 5\% of the remaining user accounts for each attack scenario, and label them as \textit{victims}. Finally, we stage the attack by replacing 5 utterances in each victim account with the attackers' utterances. 

To formulate the attack, consider the output feature vectors of the voice authentication system from a 512-dimensional space. Assume \(U_n\) with elements \(\mathbf{z}_n= \langle z_{(n,1)}, z_{(n,2)}, \ldots, z_{(n,512)} \rangle \), representing all normal accounts, and \(U_a\) representing all attacked accounts \( \mathbf{z}_a=\langle z_{(a,1)}, z_{(a,2)}, \ldots, z_{(a,512)} \rangle \). Consider \( p_n(\mathbf{z}_n | X_n) \) as the probability distribution for the normal feature vector conditioned on \( X_n \), with \( X_n \) being a subset of the input audio \( U_n \) for normal accounts. Furthermore, consider \( p_a(\mathbf{z}_a | X_a) \) as the probability distribution for the feature vectors for the compromised feature vector conditioned on \( X_a \), with \( X_a \) being a subset of the input audio \( U_a \) for compromised accounts. The hypothesis is articulated as follows \cite{10163863}:
\begin{equation}
p_n(\mathbf{z}_n | X_n) = p_a(\mathbf{z}_a | X_a), \quad \text{however,} \quad p(X_n) \neq p(X_a)
\end{equation}

Here we consider three scenarios where we train our models on 80\% of the dataset, with 5\% labeled as an attack, and evaluate on the remaining 20\%, also with 5\% labeled as an attack. We emphasize that this attack design, to our knowledge, is novel and has never been investigated before, making it the closest to a real-world poisoning attack against voice authentication systems. Addressing such an attack scenario effectively will enhance system robustness against targeted poisoning attacks.

The first scenario utilizes our proposed framework incorporating a CNN with only one block of convolutional layers (since the dataset used in this work is not large). The second scenario uses our proposed framework incorporating a CNN with two blocks of convolutional layers. Finally, the third scenario employs the proposed model by the state-of-the-art, which we refer to as the default model \cite{10163863}.

We implement our proposed framework based on the model's architecture and training process of the current state-of-the-art \cite{10163863}. To improve the methodology of \cite{10163863}, we incorporate a 5-fold cross-validation evaluation during model training on a highly imbalanced dataset, as it was not implemented in \cite{10163863}. Additionally, we integrate modifications to avoid overfitting to the data's majority class (95\%) labeled as normal. The implementation steps are as follows:

\begin{enumerate}

\item We incorporate batch normalization after each convolutional layer and both $L_1$ and $L_2$ regularization into both convolutional and dense layers to help normalize the inputs and reduce overfitting.

\item We incorporate class weights for addressing the imbalanced training data and use stratified 5-fold cross-validation in the training process to ensure that each fold is representative of the overall class distribution.

\item We conduct extensive hyper-parameter tuning to systematically explore the parameter space and determine the optimal settings.
\end{enumerate}

\begin{figure}[h]
    \centering    \includegraphics[width=3.6in]{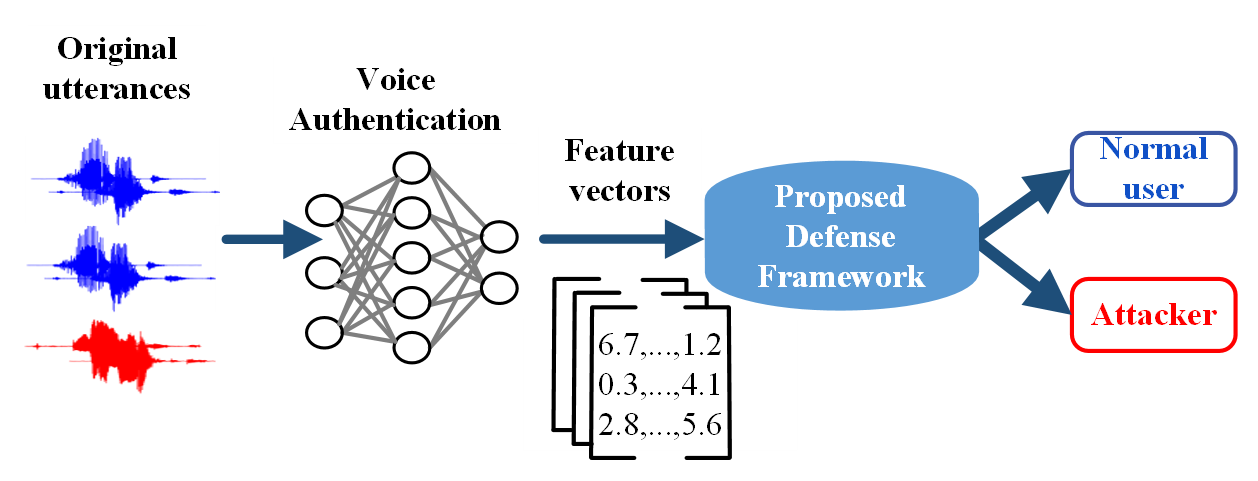}
    \caption{An overview of the authentication process flow, from the initial utterance to the final step of labeling each user as either an attacker or a legitimate user.}
    \label{fig:framework}
\end{figure}

We have pointed out significant methodological limitations in the existing works, especially the state-of-the-art \cite{10163863}, and bridged the gaps. We emphasize that incorporating the above steps and proposing new data distribution scenarios makes our methodology highly unique, novel, and realistic.

\section{PERFORMANCE EVALUATION}
For the experimental part of our study, we used the well-known  LibriSpeech dataset corpus\cite{Panayotov2015LibrispeechAA} released in 2015, composed of audiobooks in English. It contains recordings of 2,334 users, each with 10 utterances. These utterances were further trimmed to a minimum of 7 seconds in length. Moreover, we conducted our experiments using an Intel(R) Core(TM) i7-6700HQ CPU @ 2.60GHz with 12 GB RAM and a 960M model GPU. Similar to \cite{10163863}, we train Deep Speaker \cite{li2017deep} on users' data, with each user having 10 utterances.

Firstly, the users' utterances are fed to Deep Speaker, as shown in Step 1 in Figure \ref{fig:framework}. The output of Deep Speaker then goes through an embedding process similar to \cite{10163863}. Specifically, a random selection of utterances takes place pairwise from both victims and attackers.

Next, the embeddings are fed into our proposed framework, as shown in Step 2 in Figure \ref{fig:framework}. Similar to \cite{10163863}, the convolutional layers of our model contain 4 $\times$ 4 filters and a stride of 1 $\times$ 1, and 3 $\times$ 3 filters and a stride of 1 $\times$ 1 respectively, followed by max-pooling layers, dropout layers set to 20\%, and two fully connected layers in the end, totaling 12 layers. It uses a softmax cross-entropy loss function. Ultimately, the output of the CNN model is a probability value for each embedding. The KNN model aggregates the probability values for each user and labels them either as an attacker or normal, as shown in Step 3 in Figure \ref{fig:framework}.

Table \ref{tab:table_test} shows the results for each scenario on the test set. The overall accuracy, F1-score, and TP rate are highest in our proposed Scenario 1. In the first two scenarios, our proposed framework learns from both classes and generalizes to the test set with only 5\% poisoned attacks effectively. This shows our defense mechanism has been effective in addressing the real-world attack scenario, implemented on a real dataset.


\begin{table}[h]
\vspace{-10pt} 
\caption{Performance Evaluation Over All 3 Scenarios}
\centering
\footnotesize 
\begin{tabular}{|c|c|c|c|}
\hline
\textbf{Models} & \textbf{Acc.} & \textbf{Recall} & \textbf{F1} \\
\hline
Proposed Scenario 1 & 0.98 & 0.92 & 0.95 \\
\hline
Proposed Scenario 2 & 0.95 & 0.90 & 0.93 \\
\hline
Default Scenario \cite{10163863} & 0.89 & 0.55 & 0.68 \\
\hline
\end{tabular}
\label{tab:all_scenarios_performance}
\vspace{-10pt} 
\end{table}

\subsection{Performance Evaluation Over User Registration}

KNN is implemented for all three scenarios to aggregate the probability values output by the CNN model. Specifically, we run Deep Speaker 10 times as proposed by \cite{10163863}, with each instance initializing the model differently, leading to the obtaining of 10 feature vectors for each user. Then we implement the interleaved embedding on feature vectors. The embeddings are then fed to our proposed CNN model. The CNN model yields one probability value for each embedding, and each user has 10 embedding representations of their utterances. These 10 probability values become a 10-dimensional point for each user. Then, we fit a KNN model to all user points, setting K-neighbour to 11 as proposed by \cite{10163863}. This leads to a binary decision for each user: attack or normal.

\begin{table}[ht]
\caption{Evaluation results of the proposed KNN model and the default implementations (numbers in percent)}
\label{tab:comprehensive_evaluation_KNN}
\centering
\footnotesize 
\begin{tabular}{|c|c|c|c|}
\hline
\multicolumn{4}{|c|}{\textbf{Default Implementation} \cite{10163863}} \\
\hline
\textbf{Models} & \textbf{Acc.} & \textbf{Recall} & \textbf{F1} \\
\hline
Proposed Scenario 1 & 0.89 & 0.50 & 0.64  \\
\hline
Proposed Scenario 2 & 0.90 & 0.54 & 0.67 \\
\hline
Default Scenario \cite{10163863} & 0.82 & 0.36 & 0.50 \\
\hline
\multicolumn{4}{|c|}{\textbf{Proposed Implementation}} \\
\hline
Proposed Scenario 1 & 0.79 & 0.68 & 0.73 \\
\hline
Proposed Scenario 2 & 0.79 & 0.70 & 0.74 \\
\hline
Default Scenario \cite{10163863} & 0.77 & 0.48 & 0.59 \\
\hline
\end{tabular}

\end{table}

Table \ref{tab:comprehensive_evaluation_KNN} shows the results for the default KNN implementation \cite{10163863}, which indicates overfitting to the majority class and a lack of generalization to unseen data. Therefore, we implement a modified version of the KNN model to address overfitting, evident from having an accuracy above 82\% accompanied by 50\% and less recall for all three scenarios. Specifically, we incorporate the Synthetic Minority Over-sampling Technique (SMOTE), a resampling technique to address class imbalance, and stratified 5-fold cross-validation. Incorporating SMOTE and stratified cross-validation is not done in the state-of-the-art KNN model implementation \cite{10163863}. The results of our modifications are shown in Table \ref{tab:comprehensive_evaluation_KNN} under Proposed Implementation. We see increases in the recall, F1-score, and TP rate along with a reduction in FN rate.

\begin{figure}[h]
    \centering
    \includegraphics[width=4in]{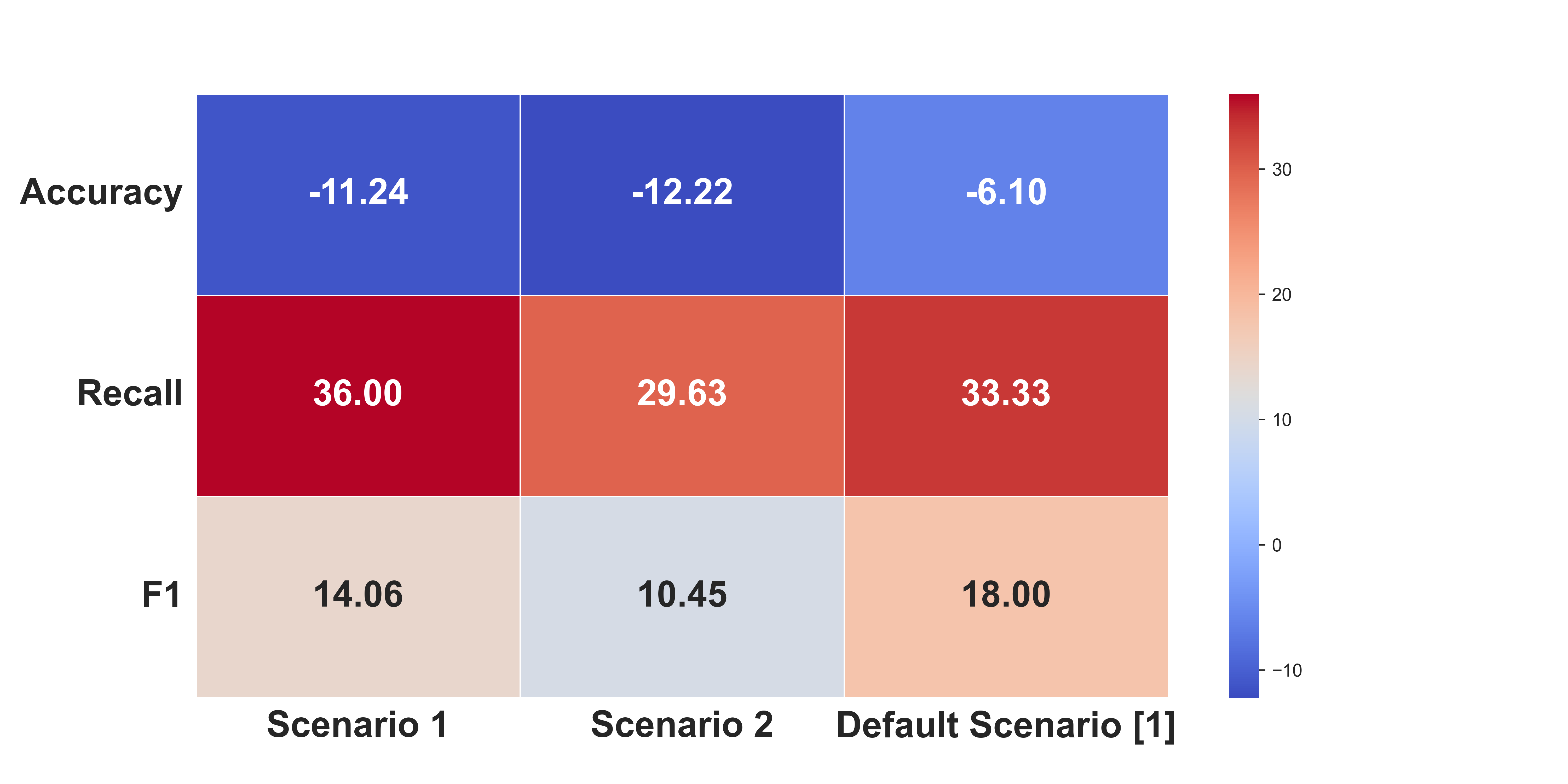}
    \caption{Heatmap Comparison of the KNN Model Default Implementation Versus Proposed Implementation.}
    \label{fig:modelPerformanceHeatmap}
\end{figure}


Figure \ref{fig:modelPerformanceHeatmap} shows the heatmap comparing the default KNN model to our proposed method, using accuracy, recall, and F1-score for Scenario 1, Scenario 2, and the Default Scenario. Despite a reduction in accuracy (-11.24\%, -12.22\%, -6.10\%), the recall (36.00\%, 29.63\%, 33.33\%) and F1-score (14.06\%, 10.45\%, 18.00\%) indicate substantial improvements in the model's ability to detect true positives and its overall detection capability. These enhancements, particularly in recall and F1-score, highlight the improved performance of our proposed model in identifying compromised accounts when compared to the Default Scenario \cite{10163863}.

\subsection{Performance Evaluation Over Different Attack Scenarios (0.1\%, 1\%, 5\%, 10\%)}
We consider different percentages of attackers staging the poisoning attack on the system at the same time. It is worth noting that the state-of-the-art \cite{10163863} considered only 10\% of users labeled as attackers. There are two main reasons to investigate the influence of having different percentages of attacker accounts. Firstly, fewer attacker accounts excluded from the data leads to more data for the training and testing of the model. Moreover, having fewer attackers leads to fewer repetitions of replaced utterances in victim accounts. This might affect the model's learning process considerably. Secondly, having fewer attackers in our attack design is closer to a real-world attack scenario, in which only a few attackers stage the poisoning attack on the system at the same time.

\begin{table}[h]
\caption{Comparison of Default and Proposed Implementations of Defense Framework for Different Attack Scenarios}
\centering
\footnotesize 
\begin{tabular}{|c|c|c|c|c|c|c|c|}
\hline
\multicolumn{4}{|c|}{\textbf{Default Implementation \cite{10163863}}} & \multicolumn{3}{|c|}{\textbf{Proposed Implementation}} \\
\hline
\textbf{Attack} & \textbf{Acc.} & \textbf{Recall} & \textbf{F1} & \textbf{Acc.} & \textbf{Recall} & \textbf{F1} \\
\hline
\textbf{0.1\%} & 1.00 & 0.99 & 0.99 & 1.00 & 0.99 & 0.99 \\
\hline
\textbf{1\%} & 0.99 & 0.84 & 0.91  & 0.95 & 0.92 & 0.94 \\
\hline
\textbf{5\%} & 0.96 & 0.60 & 0.74  & 0.82 & 0.68 & 0.74 \\
\hline
\textbf{10\%} & 0.89 & 0.50 & 0.64  & 0.79 & 0.68 & 0.74 \\
\hline
\end{tabular}
\label{tab:combined_evaluation_KNN}
\end{table}

Table \ref{tab:combined_evaluation_KNN} shows that as the number of attackers increases, accuracy and recall drop, with accuracy falling from 1.00 to 0.89 in the Default Implementation \cite{10163863} and from 0.95 to 0.79 in the Proposed Implementation, and recall from 0.99 to 0.50 and from 0.92 to 0.68, respectively. The F1-score also decreases, indicating a compromised balance of precision and recall. Conversely, FN and FP rates rise, with FN rates rising from 0.009 to 0.50 in the Default \cite{10163863} and from 0.009 to 0.30 in the Proposed Implementation. FP rates show a slight increase, notably from 0.022 to 0.01 in the Proposed Implementation. This trend suggests the model struggles more with accurate labeling as attackers' utterances diversify, leading to higher misclassification rates.

\begin{figure}[h]
    \centering    \includegraphics[width=4.2in]{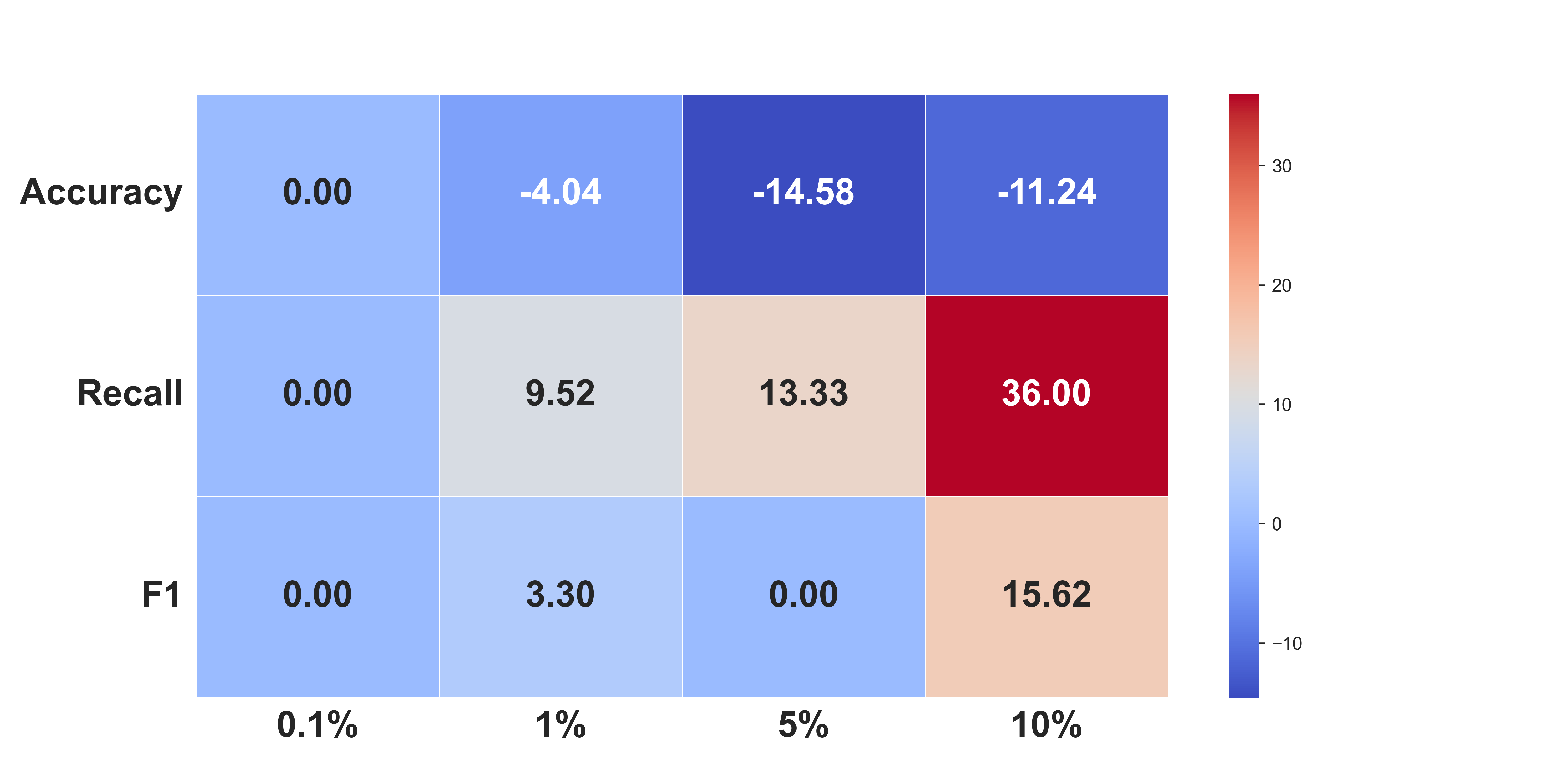}
    \caption{Heatmap Comparison of Default Defense Framework Versus Proposed Defense Framework for Different Attack Scenarios.}
    \label{fig:heat map comp knn}
\end{figure}


Figure \ref{fig:heat map comp knn} shows the heatmap comparing the default defense framework versus our proposed defense framework implementations, using accuracy, recall, and F1-score for different attack scenarios. Despite a reduction in accuracy (-11.24\%, -12.22\%, -6.10\%), the recall (36.00\%, 29.63\%, 33.33\%) and F1-score (14.06\%, 10.45\%, 18.00\%) indicate substantial improvements in the model's ability to detect true positives and its overall detection capability. These enhancements, particularly in recall and F1-score, highlight the improved performance of our proposed model in identifying compromised accounts when compared to the Default Scenario \cite{10163863}.


\section{CONCLUSION} 
 We introduced a novel approach to enhance the detection of targeted data poisoning attacks in voice authentication systems. The approach demonstrates both robustness and generalizability and it is particularly effective when the number of compromised samples is extremely low. In the future, we will explore key deployment challenges, such as the frequency of new user registrations and the corresponding need for model retraining, to understand their overall impact on the approach's performance.

\bibliographystyle{unsrt}
\bibliography{reference}

\begin{thebibliography}{10}

\bibitem{10163863}
Ke~Li, Cameron Baird, and Dan Lin.
\newblock Defend data poisoning attacks on voice authentication.
\newblock {\em IEEE Transactions on Dependable and Secure Computing}, pages 1--16, 2023.

\bibitem{Melnik2019Voice}
S.~Melnik and N.~Smirnov.
\newblock Voice authentication system for cloud network.
\newblock {\em 2019 Systems of Signals Generating and Processing in the Field of on Board Communications}, pages 1--4, 2019.

\bibitem{Paudice2018Detection}
Andrea Paudice, Luis Muñoz-González, A.~György, and Emil~C. Lupu.
\newblock Detection of adversarial training examples in poisoning attacks through anomaly detection.
\newblock {\em ArXiv}, abs/1802.03041, 2018.

\bibitem{Xu2019Adversarial}
Han Xu, Yao Ma, Haochen Liu, Debayan Deb, Hui Liu, Jiliang Tang, and Anil~K. Jain.
\newblock Adversarial attacks and defenses in images, graphs and text: A review.
\newblock {\em International Journal of Automation and Computing}, 17:151 -- 178, 2019.

\bibitem{Borgnia2020StrongDA}
Eitan Borgnia, Valeriia Cherepanova, Liam~H. Fowl, Amin Ghiasi, Jonas Geiping, Micah Goldblum, Tom Goldstein, and Arjun Gupta.
\newblock Strong data augmentation sanitizes poisoning and backdoor attacks without an accuracy tradeoff.
\newblock {\em ICASSP 2021 - 2021 IEEE International Conference on Acoustics, Speech and Signal Processing (ICASSP)}, pages 3855--3859, 2020.

\bibitem{9382920}
Dalton Cole, Sara Newman, and Dan Lin.
\newblock A new facial authentication pitfall and remedy in web services.
\newblock {\em IEEE Transactions on Dependable and Secure Computing}, 19(4):2635--2647, 2022.

\bibitem{diakonikolas2019sever}
Ilias Diakonikolas, Gautam Kamath, Daniel~M. Kane, Jerry Li, Jacob Steinhardt, and Alistair Stewart.
\newblock Sever: A robust meta-algorithm for stochastic optimization, 2019.

\bibitem{10402044}
Zhaowei Wang, Weicheng Liu, and Hui-Ming Wang.
\newblock Wireless universal adversarial attack and defense for deep learning-based modulation classification.
\newblock {\em IEEE Communications Letters}, 28(3):582--586, 2024.

\bibitem{Panayotov2015LibrispeechAA}
Vassil Panayotov, Guoguo Chen, Daniel Povey, and Sanjeev Khudanpur.
\newblock Librispeech: An asr corpus based on public domain audio books.
\newblock {\em 2015 IEEE International Conference on Acoustics, Speech and Signal Processing (ICASSP)}, pages 5206--5210, 2015.

\bibitem{li2017deep}
Chao Li, Xiaokong Ma, Bing Jiang, Xiangang Li, Xuewei Zhang, Xiao Liu, Ying Cao, Ajay Kannan, and Zhenyao Zhu.
\newblock Deep speaker: an end-to-end neural speaker embedding system, 2017.

\end{thebibliography}

\clearpage

\vfill

\end{document}